\documentclass[
    ,final            
  ]
  {aipproc}

\layoutstyle{6x9}
                                                            
\usepackage{graphicx}
\usepackage[dvips]{color}
\usepackage{colordvi}
\usepackage{amsfonts}
\usepackage{amssymb}
\usepackage{amsmath}
\usepackage{graphics}
\usepackage{pstricks}
\usepackage{bm}
\usepackage{natbib}
\newcommand{\beq}{\begin{equation}}
\newcommand{\eeq}{\end{equation}}
\newcommand{\be}{\begin{eqnarray}}
\newcommand{\ee}{\end{eqnarray}}

\begin{document}

\title{Unified description of equation of state and transport properties 
of nuclear matter}


\author{Omar Benhar$^{1,2}$, Nicola Farina$^{1,2}$, Salvatore Fiorilla$^{2}$ and 
Marco Valli$^{1,2}$}{
  address={$^1$INFN, Sezione di Roma \\ 
$^2$Dipartimento di Fisica, Universit\`a ``La Sapienza'' \\
I-00185 Roma, Italy}
}

\begin{abstract}
Correlated basis function perturbation theory and the
formalism of cluster expansions have been recently employed to
obtain an effective interaction from a state-of-the-art nucleon nucleon
potential model. The approach based on the effective interaction allows
for a consistent description of the nuclear matter ground state and
nucleon-nucleon scattering in the nuclear medium. This paper reports the
the results of numerical
calculations of different properties of nuclear and neutron matter, including
the equation of state and the shear viscosity and thermal conductivity 
transport coefficients, carried out using
the effective interaction.
\end{abstract}

\maketitle


\section{Introduction}

The theoretical description of neutron stars 
requires a quantitative understanding of both equilibrium and non equilibrium 
properties of cold nuclear matter at high density.
While equilibrium properties, e.g. the equation of state (EOS), are generally obtained 
from realistic dynamical models, strongly constrained from nuclear systematics and
nucleon-nucleon scattering data, the studies of non equilibrium behavior, based on the 
solution of the linearized Boltzmann equation, often resort to oversimplified 
models of the nucleon-nucleon (NN) interaction. 

It has been recently suggested \cite{shannon,BV:2007,exoct} that the many body 
formalism
based on correlated wave functions and cluster expansion techniques provides an
ideal framework for the unified treatment of a variety of nuclear matter properties, 
including the EOS, the dynamic response, the spin susceptibility and the transport 
coefficients. The main 
element of the approach of Ref. \cite{shannon,BV:2007,exoct} is a well behaved effective 
interaction, derived from a state-of-the-art model of nuclear dynamics, suitable for 
use in standard perturbation theory.

In this paper, after reviewing the underlying assumptions of nonrelativistic nuclear many 
body theory (NMBT), we briefly discuss the derivation of the effective interaction 
within the Correlated Basis Function (CBF) approach, and focus on
its application to the description of NN scattering in the nuclear medium.
The results of numerical calculations of the shear viscosity
and thermal conductivity of pure neutron matter, carried out using the 
in-medium NN cross section within the framework of Landau theory of normal 
Fermi liquids, are also reported.
 
\section{Formalism}

\subsection{The paradigm of Nuclear Many Body theory}

Within Nuclear Many Body Theory (NMBT), nuclear matter is viewed as a collection of 
pointlike protons and neutrons, 
whose dynamics are described by the hamiltonian
\begin{equation}
H = \sum_{i} \frac{ {\bf p}_i^2 }{ 2 m } + \sum_{j>i} v_{ij}
+ \ldots \ ,
\label{nucl:ham}
\end{equation}
$m$ and ${\bf p}_i$ being the nucleon mass and the momentum of the i-th
particle, respectively. The NN potential $v_{ij}$
reduces to the well known Yukawa one-pion exchange potential at large
distances, while its short and intermediate range behavior is determined by fitting
the available experimental data on the two-nucleon system (deuteron
properties and $\sim$ 4000 NN scattering phase shitfs). 
The state-of-the art parametrization of Wiringa {\em et al} \cite{av18}, generally 
referred to as Argonne $v_{18}$, is written in the form
\beq
v_{ij}=\sum_{n=1}^{18} v^{n}(r_{ij}) O^{n}_{ij} \ ,
\label{av18:1}
\eeq
where
\beq
O^{n \leq 6}_{ij} = [1, (\bm{\sigma}_{i}\cdot\bm{\sigma}_{j}), S_{ij}]
\otimes[1,(\bm{\tau}_{i}\cdot\bm{\tau}_{j})] \ ,
\label{av18:2}
\eeq
$\bm{\sigma}_{i}$ and $\bm{\tau}_{i}$ are Pauli matrices acting in spin and isospin 
space, respectively, and
\beq
S_{ij}=\frac{3}{r_{ij}^2}
(\bm{\sigma}_{i}\cdot{\bf r}_{ij}) (\bm{\sigma}_{j}\cdot{\bf r}_{ij})
 - (\bm{\sigma}_{i}\cdot\bm{\sigma}_{j}) \ .
\eeq
The operators with $n \leq 4$ account for the dependence of the NN force on 
the total spin and isospin of the interacting pair, while $S_{ij}$ 
produces the non central interaction responsible for the nonvanishing electric 
quadrupole moment of the deuteron. Inclusion of the the operators with $n \leq 6$ is
needed to obtain a reasonable description of the two-nucleon bound state.
The operators corresponding to $n=7,\ldots,14$ are associated with the
non static components of the NN interaction, while those
corresponding  to $p=15,\ldots,18$ account for charge symmetry violations.
The minimal set required to describe NN scattering in $S$ and $P$
states (i.e. states of relative angular momentum $L=0, 1$) consists of the 
operators corresponding to $n \leq 8$, with 
$O^{7,8}_{ij}={{\bf L}\cdot {\bf S}}\otimes[1,(\bm{\tau}_{i}\cdot\bm{\tau}_{j})]$.


The ellipses in Eq.(\ref{nucl:ham}) refer to the presence of interactions involving 
more than two nucleons. It is long known that the inclusion of a three-nucleon potential 
$V_{ijk}$ is needed to reproduce the observed binding energies of the three-nucleon system,
as well as the equilibrium density of symmetric nuclear matter \cite{uix}.
Note that, while at nuclear density 
$\langle \sum_{k>j>i} V_{ijk} \rangle \ll \langle \sum_{j>i} v_{ij} \rangle$, 
in the density regime relevant to the description of the neutron-star core 
many body forces are expected to provide a sizable contribution to the energy.

\subsection{Correlated Basis Functions}

Due to the presence of a strongly repulsive core, the matrix elements of the 
NN interaction between eigenstates of the non interacting system (Fermi gas (FG) states 
in nuclear matter) are large. As a consequence, standard 
perturbation theory is not applicable. This difficulty can be overcome 
carrying out a resummation of the perturbative expansion, 
leading to the replacement of the bare interaction with the well behaved
G-matrix (for a recent review see, e.g., Ref. \cite{marcello1}).
An alternative approach, originally developed to study strongly correlated 
quantum liquids \cite{feenberg},
is based on the replacement of the FG basis with a basis of {\em correlated}
states, embodying the nonperturbative effects arising from the short range
behavior of the NN force.
 
The {\em correlated} states of nuclear matter are obtained from FG states through 
the transformation
\begin{equation}
|n\rangle = F |n_{FG}\rangle \ ,
\end{equation}
where the operator $F$, carrying the correlation structure induced by the NN
interaction, is written in the form
\begin{equation}
F=\mathcal{S}\prod_{ij} f_{ij} \  ,
\end{equation}
$\mathcal{S}$ being the symmetrization operator. The two-body correlation functions $f_{ij}$,
whose operatorial structure reflects the complexity of the NN potential, read
\beq
f_{ij}=\sum_{n=1}^6 f^{n}(r_{ij}) O^{n}_{ij} \ ,
\label{def:corrf}
\eeq
with the $O^n_{ij}$ given by Eq.(\ref{av18:2}).

The radial functions $f^{n}(r_{ij})$ are obtained from functional minimization 
of the expectation value of the nuclear hamiltonian (\ref{nucl:ham}) in the 
correlated ground state \cite{akmal}. 

\subsection{The effective interaction}

The effective interaction $v{_{\rm eff}}$ is defined by the relation \cite{shannon}
\beq
\label{def:veff}
\langle H \rangle = \frac{\langle 0 | H | 0 \rangle}{\langle 0 | 0 \rangle} =
\langle 0_{FG} | K + v_{{\rm eff}}| 0_{FG} \rangle  \ ,
\eeq
where $H$ is the full nuclear hamiltonian of Eq.(\ref{nucl:ham}) and $K$ is the kinetic
energy operator.

In order to include interactions involving more than two nucleons, 
we have followed the approach originally proposed in Ref. \cite{LagPan}, in which the 
main effect of 
three- and many-body forces is taken into account through a density dependent modification 
of the NN potential $v_{ij}$ at intermediate range. Moreover, in view of the weak model 
dependence of our results \cite{BV:2007}, the full $v_{18}$ potential has 
been replaced by 
its reduced form $v^\prime_{8}$ \cite{V8P}, in which only the components with $n\leq8$ are 
retained. 

The cluster expansion technique \cite{jwc} allows one to rewrite $\langle H \rangle$ in the 
form
\beq
\langle H \rangle = K_{FG} + \sum_{n\geq2} (\Delta E)_n \ ,
\eeq
where $K_{FG}$ is the Fermi gas kinetic energy and $(\Delta E)_n$ is the 
contribution to the energy arising from $n$-nucleon clusters. 

To obtain $v_{{\rm eff}}$ from Eq.(\ref{def:veff}), $\langle H \rangle$
is evaluated at the two-body level of the cluster expansion \cite{shannon}.
The resulting effective interaction reads
\beq
\label{veff:2B}
v_{{\rm eff}} = \sum_{i < j} f_{ij}^\dagger \left[ -\frac{1}{m} (\nabla^2 f_{ij}) 
 - \frac{2}{m} (\bm{\nabla} f_{ij}) \cdot \bm{\nabla} + v_{ij}f_{ij} \right]  \ .
\eeq

For any given density, the radial functions $f^n(r_{ij})$ of Eq.(\ref{def:corrf}) are solutions
of a set of Euler-Lagrange equations satisfying the boundary conditions $f^1(r_{ij} \geq d) = 1$,
$f^n(r_{ij}~\geq~d)~=~0$, for $n =2, 3$ and $4$, and  $f^n(r_{ij} \geq d_t) = 0$,
for $i = 5, 6$ (see, e.g., Ref. \cite{LagPan}).
Note that, as the non static terms in Eq.(\ref{veff:2B}) yield a negligibly small
contribution to the ground state energy \cite{shannon}, they have been neglected 
altogether. 

\begin{figure}[htb]
\includegraphics[height=.35\textheight]{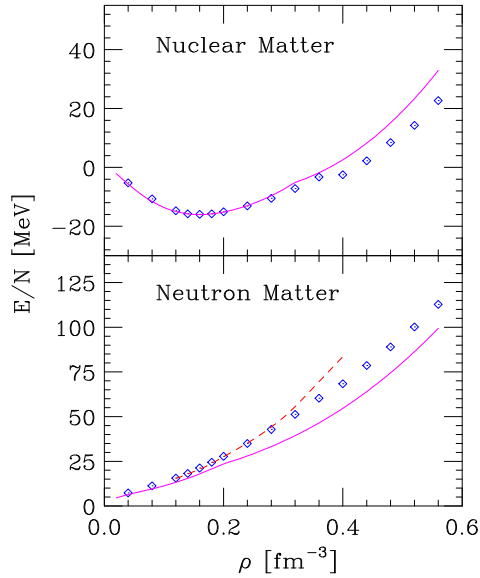}
\caption{ 
Energy per particle of symmetric nuclear matter (upper panel) and pure neutron matter
(lower panel). The diamonds represent the results obtained using the effective interaction
discussed in the text in first order perturbation theory with the FG basis,
whereas the solid lines correspond to the results of Akmal, Pandharipande 
and Ravenhall \cite{akmal}. The dashed line of the lower panel represents the results
of the AFDMC approach or Ref. \cite{AFDMC}. \label{energy_new} }
\end{figure}

The effective interaction of Eq.(\ref{veff:2B}) was tested by computing the energy
per particle of symmetric nuclear matter and pure neutron matter in first order perturbation
theory using the FG basis. In Fig. \ref{energy_new} our results are compared to those of
Refs. \cite{akmal} and \cite{AFDMC}.
The calculations of Ref. \cite{akmal} (solid lines) have been carried out using a 
variational approach based on the FHNC-SOC formalism, with a hamiltonian including the 
Argonne $v_{18}$ NN potential and the Urbana IX three-body potential \cite{uix}. 
The results of Ref. \cite{AFDMC} (dashed line of the lower panel) have been obtained 
using the $v_8^\prime$ and the same three-body potential within the framework of the 
Auxiliary Field Diffusion Monte Carlo (AFDMC) approach.

The results of Fig. \ref{energy_new} show that the effective interaction
provides a fairly reasonable description of the EOS over a broad density range. Note that
empirical equilibrium properties of symmetric nuclear matter are accounted for without
including the somewhat {\em ad hoc} density dependent correction of Ref. \cite{akmal}.
It should also be emphasized that our approach does not involve adjustable parameters. 
The correlation ranges
$d$ and $d_t$ have been taken from Ref. \cite{akmal}, while the parameters entering
the definition of the three-nucleon interaction (TNI) have been determined by
the authors of Ref. \cite{LagPan} through a fit of nuclear matter equilibrium
properties.

\subsection{Transport properties of interacting Fermi systems}

The theoretical description of transport properties of normal Fermi liquids
is based on Landau theory \cite{baym-pethick}. 
Working within this framework and including the leading
term in the low-temperature expansion, Abrikosov and Khalatnikov (AK) \cite{ak} obtained
approximate expressions for the shear viscosity and the thermal conductivity.
Let us consider viscosity, as an example. The AK result reads
\beq
\label{eta_AK}
\eta_{AK} = \frac{1}{5}\rho m^\star v^2_F \tau \,\frac{2}{\pi^2(1-\lambda_\eta)} \ ,
\eeq
where $\rho$ is the density, $v_F$ is the Fermi velocity and $m^\star$ and $\tau$ denote
the quasiparticle effective mass and lifetime, respectively. The latter can be
written in terms of the angle-averaged scattering probability, $\langle W \rangle$,
according to
\beq
\label{tau_AK}
\tau T^2 = \frac{8\pi^4}{{m^*}^3}\ \frac{1}{\langle W \rangle} \ ,
\eeq
where $T$ is the temperature and
\beq
\label{Wavg}
\langle W \rangle = \int \frac{d\Omega}{2\pi}\ \frac{W(\theta,\phi)}
{\cos{(\theta/2)}} \ .
\eeq
Note that the scattering process involves quasiparticles on the Fermi surface.
As a consequence, for any given density $\rho$, $W$ depends only on
the angular variables $\theta$ and $\phi$, the magnitude of all quasiparticle momenta
being equal to the Fermi momentum $p_F=(3\pi^2 \rho)^{1/3}$.
Finally, the quantity $\lambda_\eta$ appearing in Eq.(\ref{eta_AK}) is defined as
\beq
\lambda_\eta = \frac{\langle W [ 1-3\sin^4{(\theta/2)}\sin^2{\phi} ] \rangle}
{\langle W \rangle} \ .
\eeq
The exact solution of the equation derived in Ref. \cite{ak}, obtained by
Brooker and Sykes \cite{sb1}, reads
\beq
\nonumber
\eta = \eta_{AK} \ \frac{1-\lambda_\eta}{4} 
 \sum_{k=0}^\infty \frac{4k+3}{(k+1)(2k+1)[(k+1)(2k+1)-\lambda_\eta]} \ ,
\label{eta_sb}
\eeq
the size of the correction with respect to the result of Eq.(\ref{eta_AK}) being
$0.750 < (\eta/\eta_{AK}) < 0.925$.

Eqs.(\ref{eta_AK})-(\ref{eta_sb}) show that the key element in the determination
of the viscosity
is the in-medium NN scattering cross section. In Ref. \cite{panpiep}, the relation
between NN scattering in vacuum and in nuclear matter has been analyzed under
the assumption that the nuclear medium mainly affects the flux of incoming
particles and the phase space available to the final state particles, while leaving
the transition probability unchanged. Within this picture $W(\theta,\phi)$ can
be extracted from the NN scattering cross section measured
in free space, $(d\sigma/d\Omega)_{\rm{vac}}$, according to
\beq
\label{Wfree}
W(\theta,\phi) = \frac{16 \pi^2}{{m^\star}^2}
\left( \frac{d\sigma}{d\Omega} \right)_{{\rm vac}}  \,
\eeq
where $m^\star$ is the nucleon effective mass and $\theta$ and $\phi$ are related to
the kinematical variables in the center of mass frame
through $E_{cm} = p_F^2(1-\cos \theta)/(2m)$, $\theta_{cm} = \phi$ \cite{flowi2}.
                                                                                           
The above procedure has been followed in Ref. \cite{haensel}, whose authors
have used the available tables of vacuum cross sections obtained from partial wave
analysis \cite{SAID}. In order to compare with the results of Ref. \cite{haensel},
and gauge the model dependence of our results, we have carried out a calculation 
of the viscosity using Eqs.(\ref{eta_AK})-(\ref{Wfree})
and the free space neutron-neutron cross section obtained from the 
Argonne $v_{18}$ and $v^\prime_{8}$ potentials. 

\begin{figure}[ht]
  \includegraphics[height=.25\textheight]{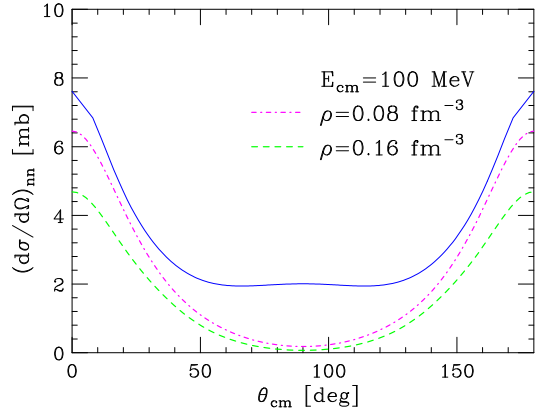}
  \caption{ Differential neutron-neutron scattering cross section at $E_{cm}=100$ MeV,
as a function of the scattering angle in the center of mass frame. Solid line: cross
section in vacuum, calculated with the $v_8^\prime$ potential \cite{V8P}.
Dot-dash line: medium modified cross section
obtained from the effective interaction described in the text at $\rho = 0.08$ fm$^{-3}$.
Dashed line: same as the dot-dash line, but for $\rho = 0.16$ fm$^{-3}$. \label{dsigma} }
\end{figure}

\begin{figure}[hbt]
  \includegraphics[height=.35\textheight]{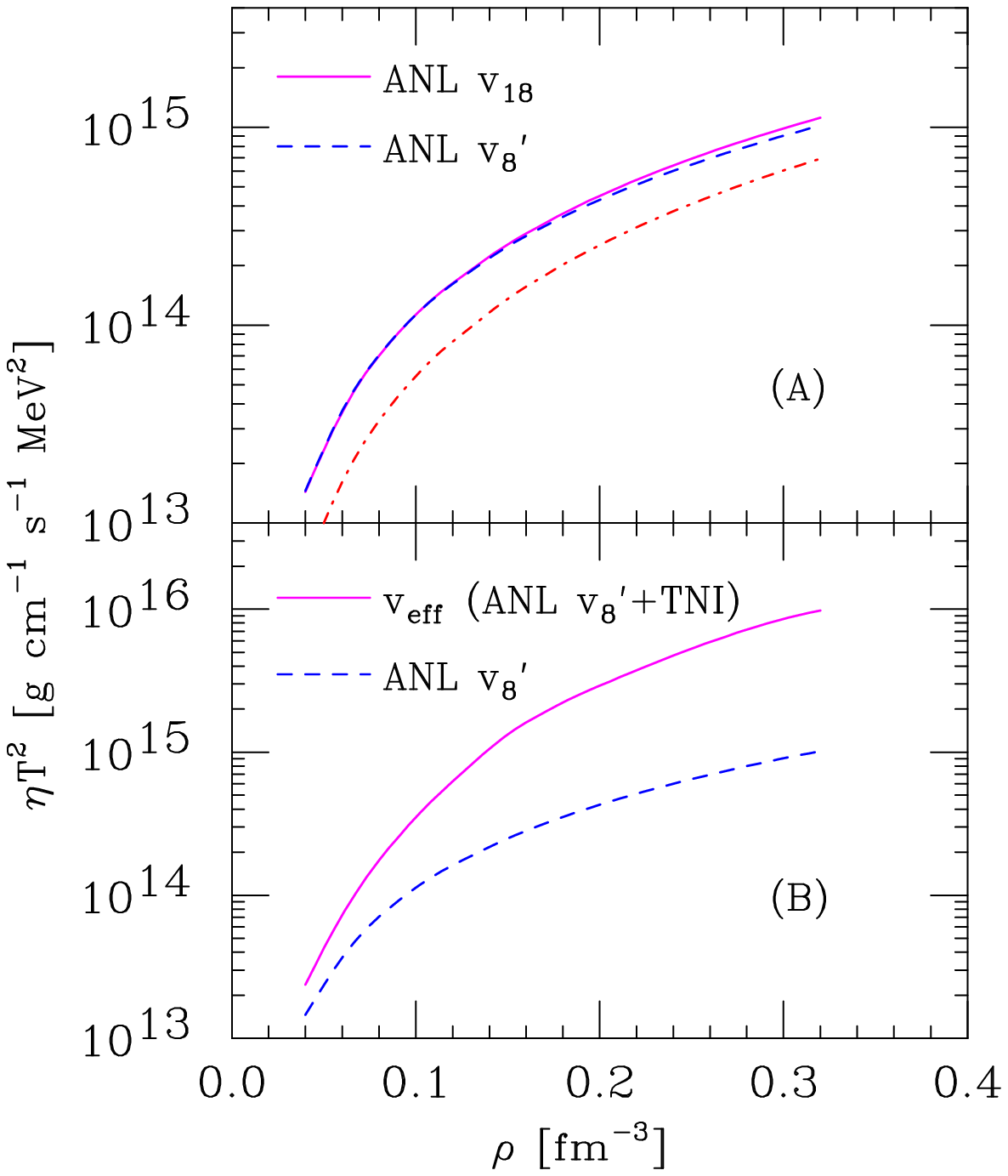}
  \includegraphics[height=.25\textheight]{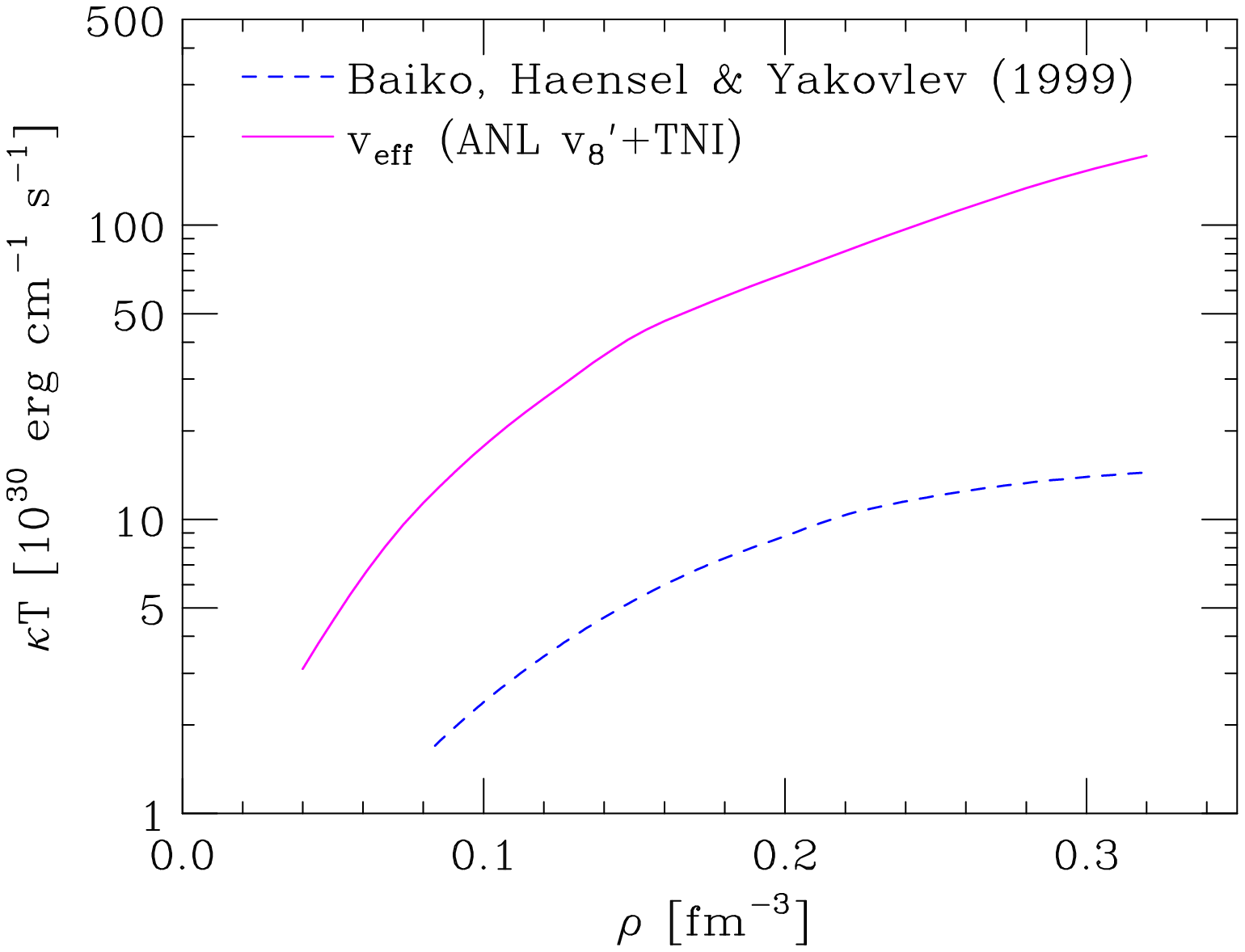}
  \caption{Left Panel: neutron matter $\eta T^2$ as a function of density.
(A). Solid line: results obtained from
Eqs.(\ref{eta_AK})-(\ref{Wfree}) using the Argonne $v_{18}$ potential and
$m^\star$ computed from the effective interaction described in the text.
Dot-dash line: results obtained from Eqs.(43) and (46) of Ref. \cite{haensel}
using the same $m^\star$. Dashed line: same as the solid line, but with the
Argonne $v_{18}$ replaced by its reduced form $v^\prime_{8}$.
(B). Solid line: results obtained using the effective interaction
described in the text. Dashed line: $\eta T^2$ obtained from the free space cross section
corresponding to the $v_8^\prime$ potential.  Right panel:
neutron matter $\kappa T$ as a function of density. Solid line: calculation
carried out using the approach of Ref. \cite{BV:2007};
dashed line: results of Baiko, H\"ansel \& Yakovlev \cite{BHY}.
\label{eta} }
\end{figure}

In the upper left panel of Fig. \ref{eta}, we show the quantity $\eta T^2$ as a
function of density. Our results are represented by the solid line, while the dot-dash 
line corresponds to the results obtained from Eqs.(43) and (46) of Ref. \cite{haensel}
using the same effective masses, computed from the effective interaction discussed above.
The differences between the two curves are likely to be ascribed to the
to the extrapolation needed to determine the cross sections
at small angles within the approach of Ref. \cite{haensel}.

To improve upon the approximation of Eq.(\ref{Wfree}) and
include the effects of medium modifications of the NN scattering amplitude, we
have replaced the bare NN potential with the CBF effective interaction.

Knowing the effective interaction, the in-medium scattering probability can be
readily obtained from Fermi's golden rule. The corresponding cross section
at momentum transfer ${\bf q}$ reads
\beq
\frac{d\sigma}{d\Omega} = \frac{{m^\star}^2}{16 \pi^2}
\ |\hat{v}_{eff}({\bf q})|^2 \ ,
\label{sigma:medium}
\eeq
$\hat{v}_{eff}$ being the Fourier transform of the effective potential.
The effective mass can also be extracted from the quasiparticle energies computed
in Hartree-Fock approximation. For symmetric nuclear matter at equilibrium, we
find $m^\star(p_F)/m = 0.65$, in close agreement with the lowest order CBF result
of Ref. \cite{FFP}.

In Fig. \ref{dsigma} the in-medium neutron-neutron cross section at $E_{cm}=100$ MeV
obtained from the effective potential, with $\rho = \rho_0$ and $\rho_0/2$ 
($\rho_0=$0.16 fm$^{-3}$ is the equilibrium density of nuclear matter), is compared
to the corresponding free space result. As expected, screening of the bare
interaction leads to an appreciable suppression of the scattering cross section.

Replacing the cross section in vacuum with the one defined in Eq.(\ref{sigma:medium}),
 the medium modified scattering probability can be obtained from Eq.(\ref{Wfree}).
The resulting $W(\theta,\phi)$ can then be used to calculate $\eta T^2$ from
Eqs.(\ref{eta_AK})-(\ref{eta_sb}). Similar expression are found for the thermal
conductivity.

The effect of using the medium modified cross section is illustrated in the 
lower left panel of Fig. \ref{eta}.
Comparison between the solid and dashed lines shows that inclusion of medium
modifications leads to a large increase of the viscosity, ranging between a factor
$\sim$ 2.5 at half nuclear matter density to a factor $\sim$ 10
 at $\rho = 2 \rho_0$. Such an increase is likely to produce appreciable effects on
the damping of neutron-star oscillations associated with emission of gravitational 
waves \cite{lindblom}.

The right panel of figure \ref{eta} shows that medium modification of NN scattering 
also produce a
large effect on the thermal conductivity, $\kappa$. The comparison between our results and 
those of Ref. \cite{BHY}, obtained from the free space NN cross sections, shows
a difference in $\kappa T$ exceeding one order of magnitude at $\rho = 2 \rho_0$.
The impact of this result on neutron star cooling is currently being investigated 
\cite{fiorilla}.

\section{summary}

Working within the formalism based on correlated functions and cluster expansion 
techniques, we have derived an effective interaction from a realistic
NN potential model. 
Our work improves upon the CBF effective interaction of 
Ref. \cite{shannon} in that it includes the effects of many-nucleon forces, which 
become sizable, indeed dominant, in the high density region relevant to the studies 
of neutron star properties.
The energy per nucleon of both symmetric nuclear matter and pure neutron matter, obtained
from our effective interaction model, turns out to be in fairly good agreement with
the results of highly refined many body calculations, based on similar dynamical
models. 
The emerging picture suggests that our approach captures the relevant physics,
allowing for a unified description of a number of different properties of 
neutron star matter based on standard perturbation theory in the FG basis.

%
%


\bibliographystyle{aipproc}   

\bibliography{benhar}

\IfFileExists{\jobname.bbl}{}
 {\typeout{}
  \typeout{******************************************}
  \typeout{** Please run "bibtex \jobname" to optain}
  \typeout{** the bibliography and then re-run LaTeX}
  \typeout{** twice to fix the references!}
  \typeout{******************************************}
  \typeout{}
 }

\end{document}